\documentstyle[epsfig]{elsart}

\def\nim#1#2#3  {{\em Nucl. Instr. Meth.} {\bf#1} (#2) #3.}
\def\pl#1#2#3   {{\em Phys. Lett.} {\bf#1} (#2) #3.}
\def\prl#1#2#3  {{\em Phys. Rev. Lett.} {\bf#1} (#2) #3.}

\begin{document}


\newcommand{\bc}{\begin{center}}
\newcommand{\ec}{\end{center}}

\begin{frontmatter}

\title{ Development of a high-efficiency pulsed slow positron beam for
  measurements with orthopositronium in vacuum.}

\begin{center}

\author[LMOPS]{N.~Alberola},
\author[LAPP]{T.~Anthonioz},
\author[ETH]{A.~Badertscher},
\author[LMOPS]{C.~Bas},
\author[INR]{A.S.~Belov},
\author[ETH]{P.~Crivelli},
\author[INR]{S.N.~Gninenko},
\author[INR]{N.A.~Golubev},
\author[INR]{M.M.~Kirsanov},
\author[ETH]{A.~Rubbia},
\author[LAPP]{D.~Sillou}
\end{center}

\address[LMOPS]{LMOPS, Le Bourget du Lac, CNRS,  France}
\address[LAPP]{LAPP, Annecy le Vieux, CNRS-IN2P3,  France}
\address[ETH]{Institut f\"ur Teilchenphysik, ETHZ, CH-8093  Z\"urich, Switzerland}
\address[INR]{Institute for Nuclear Research of the Russian Academy of
  Sciences, Moscow 117312, Russia}

\begin{abstract}

We have developed a high-efficiency pulsed slow positron beam for experiments with
orthopositronium in vacuum. The new pulsing scheme is based on a double-gap coaxial
buncher powered by an RF pulse of  appropriate shape.
The modulation of the positron velocity in the two gaps is used
to adjust their time-of-flight to a target.
This pulsing scheme allows to minimize non-linear aberrations in the bunching
process and to
efficiently compress positron pulses with an initial pulse duration ranging from $\sim$
300  to 50 ns into bunches of 2.3 to 0.4 ns width, respectively, with a repetition period
of 1~$\mu s$. The compression ratio achieved is $\simeq 100$, which is a factor 5 better
than has been previously obtained with slow positron beams based on a single buncher.
Requirements on the degree, to which the moderated positrons should be mono-energetic and
on the precision of the waveform generation are presented. Possible applications of the
new pulsed positron beam for measurements of thin films are discussed.

\end{abstract}

\end{frontmatter}

%








\section{Introduction}

Orthopositronium ($o-Ps$), the triplet $e^+e^-$-bound state, is a particularly
interesting system for a search for new physics \cite{work} - \cite{scalc} and for high
precision tests of QED \cite{karsh} -\cite{rich}. New experimental results on this system
are expected by improving the sensitivity of the previously developed techniques based on
the $o-Ps$ production in a low density SiO$_2$ target \cite{pc}. Other possible
experiments, such as the search for mirror type dark matter via the $o-Ps \to invisible$
decay mode \cite{bader}-\cite{fg}, precise measurements of the $o-Ps$ decay rate
\cite{ds} and others \cite{mesh}, require the production and subsequent decay of $o-Ps$
to occur in vacuum. For these experiments a  specially designed slow positron beam
operating in a pulsed mode with a repetition period of $\simeq 1~ \mu s$ has been
constructed. The construction of the beam has to compromise two main design goals
\cite{bader}: i) the time of the primary positron collection has to be comparable with
the repetition period in order to get the highest pulsing efficiency (the ratio of beam
intensities in the pulsed and DC modes) and to enhance the signal-to-noise ratio; ii) a
high beam compression factor of $\simeq 100$ has to be achieved in order to obtain a
positron pulse width of a few ns and to suppress background for tagging of $o-Ps$
production.

Various techniques to produce pulsed positron beams have been reported with the main
focus so far on material science applications, for a review, see  e.g. \cite{col,charl}.
The Munich group uses a pulsing scheme with two bunchers \cite{munich}. A saw-tooth shape
pulse is used in a pre-buncher to produce 2 ns pulses from the initial pulse of 13 ns.
Then, a  50 MHz chopper is used to eliminate unbunched positrons and to increase the
signal to background ratio. Finally, a 50 MHz RF main buncher produces longitudinal
compression of pulses from 2 ns to 150 ps (FWHM) duration. A similar pulsing scheme is
used by the Tsukuba group \cite{tsuk}. For the vacuum experiments mentioned above, this
method requires a wide time window of chopping, and accordingly, the positron collection
efficiency from an initial DC beam would become less than 1\%.

A different pulsing method has recently been considered by Oshima et al. \cite{oshima}.
The main idea is the same as for the RF method: the time-of-flight for each positron is
adjusted according to the time it arrives at the starting point of the acceleration.
However, instead of applying a sinusoidal-like RF field, a more suitable pulse shape of
the electric field, with an approximate inverse parabolic function of time, is applied to
a single gap for the positron velocity modulation \cite{hamada}. This method has been
further developed by Iijima et al. \cite{iijima} for material measurements in which the
lifetime of orthopositronium atoms is close to its vacuum value of $\simeq$ 142 ns. For
these applications it is necessary to  modify the originally proposed technique in order
to generate higher intensity positron beams by accumulating positrons over a wider time
interval, even though the bunch width becomes larger, but is still much less than the
typical measured timing intervals of $\simeq$ 100 ns. Using a high permeability buncher
core, a bunch width of 2.2 ns (FWHM) for 50 ns collection time and a repetition period of
960 ns has been achieved \cite{iijima}. One of the problems encountered is the limitation
of the voltage supplied by a post-amplifier to the buncher.

In this paper we describe a double-gap coaxial buncher powered by an RF pulse of
appropriate shape, which is produced by an arbitrary waveform generator (AWG) and by a
post-amplifier. This pulsing method allows to reduce the influence of aberrations of the
bunching pulse shape in comparison with methods using a sinusoidal RF voltage and to
achieve a compression ratio limited mainly by the intrinsic energy spread of the initial
positrons. In comparison with the one gap buncher method, see e.g. Ref. \cite{oshima},
the present scheme requires lower bunching voltage and less post-amplifier power
\footnote{More details on a pulsing system with a single-gap velocity modulation can be
found in ref. \cite{oshima,hamada}}.


The rest of this paper is  organized as follows:
The beam prototype and the new pulsing system are described in
Section~2. The description of simulations used to design the pulsing system is given in
Section~3. In Section~4 the results obtained with the pulsed beam are presented and the
requirements for the system components are discussed. Possible applications of the
developed pulsed beam for measurements in material research are discussed in Section 5. A
summary is given in Section 6.

\begin{figure}[htb!]
\hspace{-.cm}{\epsfig{file=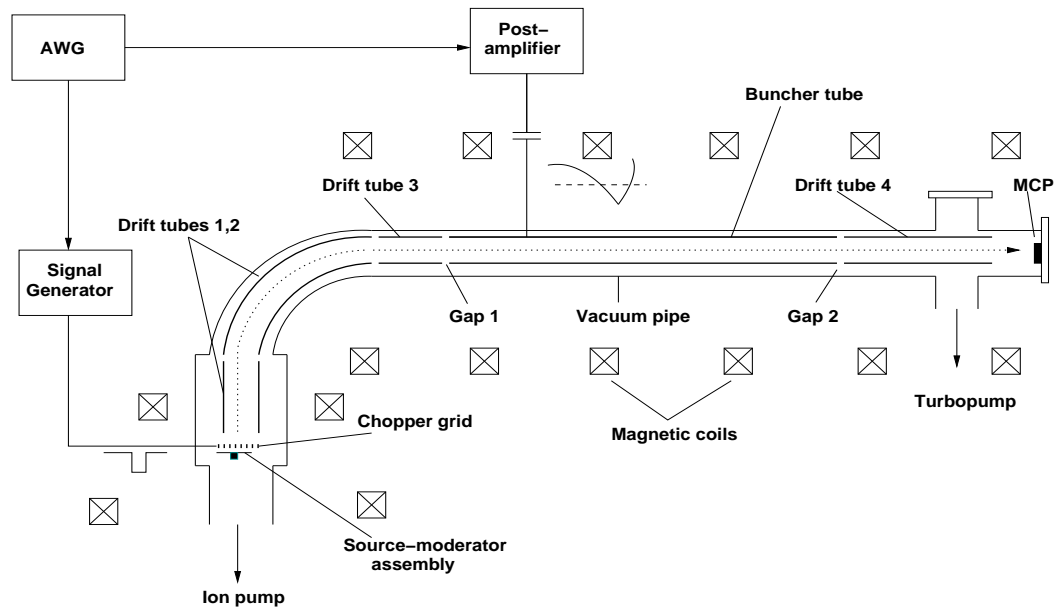,width=140mm,height=80mm,
angle=0}}
 \caption{\em Schematic illustration of the magnetically transported pulsed
positron beam.}
\label{beam}
\end{figure}

\section{The pulsed slow positron beam}

\subsection{The actual beam}

\begin{figure}[htb!]
\begin{center}
\hspace{-0.cm}{\epsfig{file=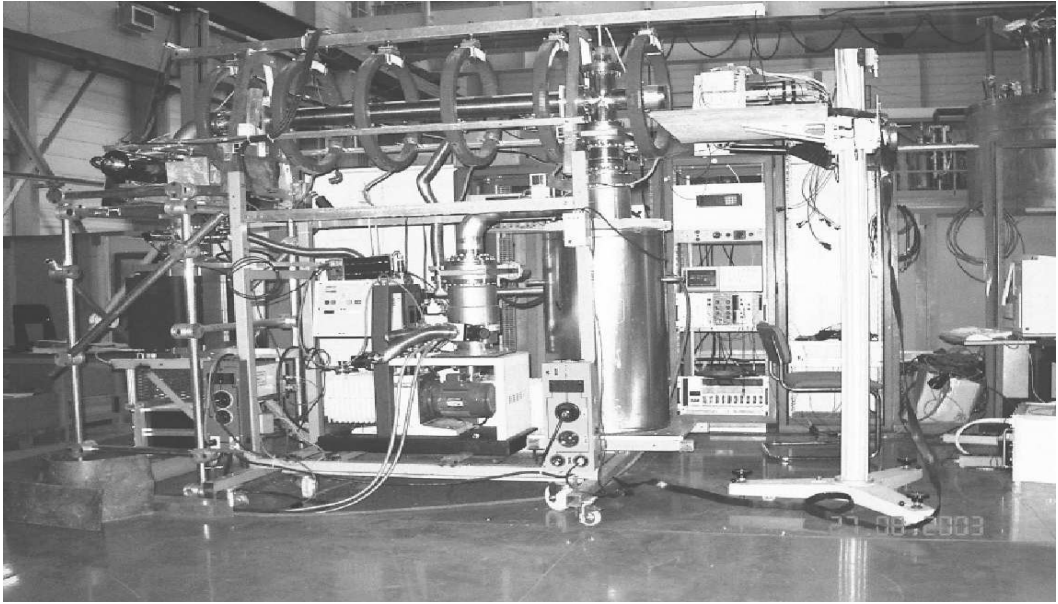,width=140mm,height=80mm,angle=0}}
\end{center}
 \caption{\em Photograph of the slow positron pulsed beam prototype.}
\label{beam_photo}
\end{figure}

The preliminary design of the present pulsed slow positron beam has been reported in
\cite{bader}. Our primary consideration is that the system should be of the magnetic
transport type because this provides the simplest way to transport a slow positron beam
from the positron source to its target \cite{orli}. Fig. \ref{beam}
and Fig. \ref{beam_photo} show the schematic
illustration of the developed pulsed slow positron beam and the
photgraph of the actual beam, respectively.  The DC slow positrons are
produced by moderating the fast positrons emitted in $\beta^+$-decays of the radioisotope
source $^{22}$Na. This source, with a relatively small activity ( $\sim 50\mu$Ci) was
prepared by bombarding a 150 $\mu$m thick foil of pure Al with a 590 MeV proton beam at
the PSI accelerator (Paul Scherrer Institute, Switzerland).

The moderator, a tungsten W(100) single crystal foil with a thickness of 1 $\mu m$, is
prepared {\em in situ} by annealing it at 2000$^o$C (see Section 4). A few eV positrons
from the moderator are accelerated to 30 eV and are separated from the fast unmoderated
positrons by a 90$^o$ curved B-field, serving as a velocity filter. The eight coils
provide a quasi-uniform longitudinal magnetic field of 70 Gauss to guide positrons down
to the target - a microchannel plate (MCP, Hamamatsu F4655-12), located at the end of the
beam line and used for positron detection. The beam energy can be varied up to a few kV
simply by applying the desired electrostatic potentials to the beam drift tubes.

\begin{figure}[htb!]
\begin{center}
  \epsfig{file=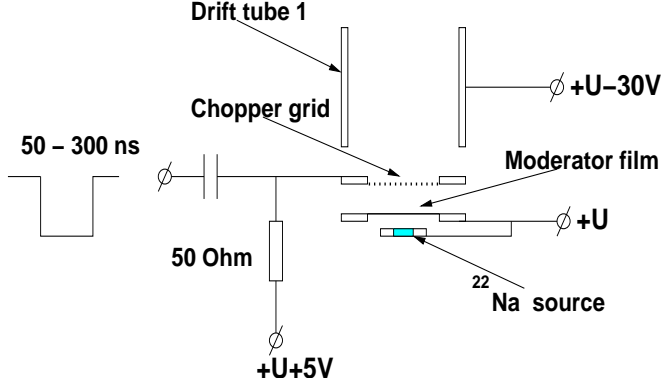, width=90mm,height=50mm}
\end{center}
 \caption{\em Schematic illustration of the source-moderator assembly.}
\label{mod_scheme}
\end{figure}
The positron pulsing section shown in Fig. \ref{beam} consists of a chopper, the drift
tubes, and a buncher. The latter is constructed in such a way that its internal tube (the
buncher electrode) forms a coaxial line of 50 $\Omega$ impedance with the external vacuum
pipe. The buncher internal tube length of 135 cm is determined by the distance-of-flight
of positrons entering the buncher during the initial pulse and by their energy. For the
initial pulse of 90 ns the repetition period can be varied from 220 ns to infinity.
Initial positron pulses with the desired duration are formed with the chopper grid placed
4 mm downstream of the moderator foil, as shown in Fig. \ref{mod_scheme}. \\
The potential difference $\Delta U=$5 V between the moderator film and the grid is used
for repelling the slow $\simeq 3$ eV positrons emitted from the moderator. When the
rectangular shaped chopper pulse of a given duration (50 - 300 ns, produced by a standard
fast-signal generator), is applied to the grid with an amplitude of about 5V, the
moderated positrons come through and are accelerated in the gap between the chopper grid
and the first drift tube.

\subsection{The pulsing scheme for the buncher}

\begin{figure}
\begin{center}
  \epsfig{file=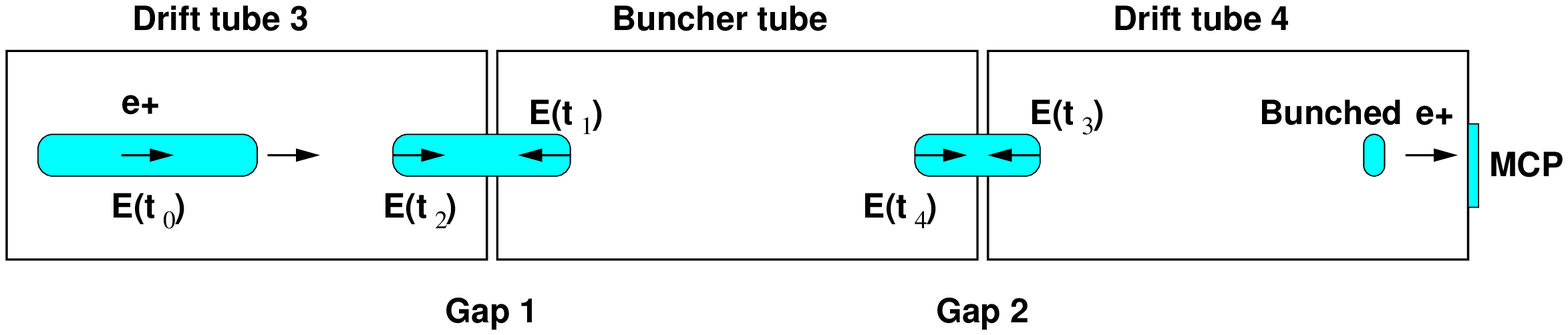,width=120mm,height=30mm}
\end{center}
 \caption{\em Principle of the positron beam compression with a double gap
   buncher: at the time $t_0$ all chopped positrons have received initial
   acceleration, at the time $t_1>t_0 (t_3>t_2)$ positrons passing the first
   (second) gap early are decelerated, while those which pass the gap at a later
time $t_2>t_1 (t_4>t_3)$ are accelerated. After this velocity modulation, positrons
arrive at the target position (MCP) as a short bunch.} \label{schema}
\end{figure}

The principle of the positron pulsing method is illustrated in Fig. \ref{schema}. A
non-linear time-varying electric field in the first gap (Gap1) between the drift tube 3
and the bunching electrode modulates the velocity of positrons in such a way, that those
(initially chopped and accelerated) positrons which arrive early are decelerated at the
gap, while those which reach the gap later are accelerated.
In the second gap (Gap2) between the buncher and drift tube 4 the same procedure is
repeated. The chopper pulse is synchronized with the AWG trigger signal such that the
chopped initial positrons passing the bunching electrode receive the correct bunching
voltage from the corresponding part of the bunching pulse. Finally, the initially chopped
positron pulse arrives at the MCP target as a bunch of a small width.

The bunching voltage of the designed shape is produced by the arbitrary waveform
generator (AWG-510, Tektronix). The AWG output pulse is then amplified by the post
amplifier 100A250A (Amplifier Research) and is applied to the bunching electrode. The
amplifier has an output power of 100 W and a bandwidth of 10 kHz - 250 MHz, allowing to
drive the 50 $\Omega$ coaxial buncher with a pulse amplitude up to $\simeq 80$ V without
saturation. The time profile of the bunched positrons at the MCP position is measured by
using the MCP signal as the TDC-START and the AWG signal as the STOP; the STOP signal is
also used to trigger the chopper pulse.

\section{Design of the pulsing system}

The simulations of the E- and B-fields, the beam transport and the velocity modulation of
positrons were performed with the GEANT4 \cite{geant} and 3D-Bfield \cite{nu} codes with
the goal to minimize the time width of the bunch and to optimize the shape of the
bunching pulse.

The optimal shape and duration of the bunching pulse was calculated for the given initial
positron pulse duration by taking into account the following criteria:

\begin{itemize}
\item the amplitude of the RF bunching pulse should be within $\pm$80 V,
\item after the velocity modulation positrons should arrive
at the target as a bunch within a time spread of $\simeq 1$ ns,
\item absence of significant non-linear distortions of the beam phase-space
at the target position
 \end{itemize}

\begin{figure}
\begin{center}
  \epsfig{file=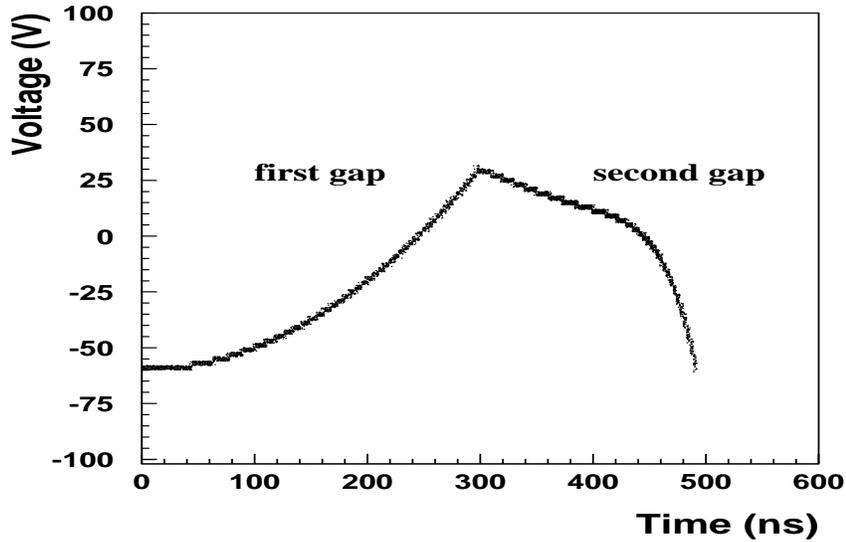,width=120mm,height=80mm}
\end{center}
 \caption{\em  The bunching voltages seen by the positrons at the
first and the second velocity modulation gaps, respectively.} \label{pulse}
\end{figure}

For the first gap and for an initial pulse of e.g. 300 ns, a parabolic time-varying
potential $V(t)\sim t^2$ changing from -60 V (decelerating part) to +30 V has been
chosen, as shown in Fig. \ref{pulse}. The time dependence of the potential at the central
electrode of the buncher at times $t>300$ ns can then be calculated, solving the
corresponding equations with an iterative procedure under the condition that positrons
arrive at the target simultaneously and that the potential at the electrode at the end of
the bunching pulse returns to its initial value -60 V. In Fig. \ref{pulse} the resulting
shape of the bunching voltage is shown for both gaps.

In reality, the bunching pulse shape differs from its ideal theoretical shape, mostly due
to the finite frequency bandwidth of the post-amplifier and due to non-ideal matching of
the coaxial buncher curcuit to 50 $\Omega$ impedance. To estimate the effect, the
response of the 100A250A amplifier was simulated according to its circuit characteristics
\cite{ra}. The residual shape is defined as $R(t) = S_{out}(t)-S_{in}(t)$, where
$S_{in}(t)$ is the input signal supplied by the AWG with unit amplitude and $S_{out}(t)$
is the amplifier output pulse calculated for unit gain.

It was found that for an initial pulse duration of 300 ns the deviation of the response
from the ideal shape shown in Fig.\ref{pulse} is not more than about $R(t)/S_{out}
\pm$1\% over the full bunching pulse duration. The simulations show, that such a signal
deviation results in a negligible distortion of the bunched positron pulse shape. The
FWHM of the corresponding distribution has been increased by less than 2\%. However, for
shorter initial pulses ($ < 100$ ns ), deviations up to about $R(t)/S_{out} \pm$5\% have
been observed. This will result in a significant degradation of the FWHM and shape of the
positron pulses. These results mean that the theoretical shape of the bunching pulse must
be reproduced within about $\pm$1\%. For an initial positron pulse duration of more than
$\simeq$ 100 ns this is achievable.

\begin{figure}[htb!]
\begin{center}
  \epsfig{file=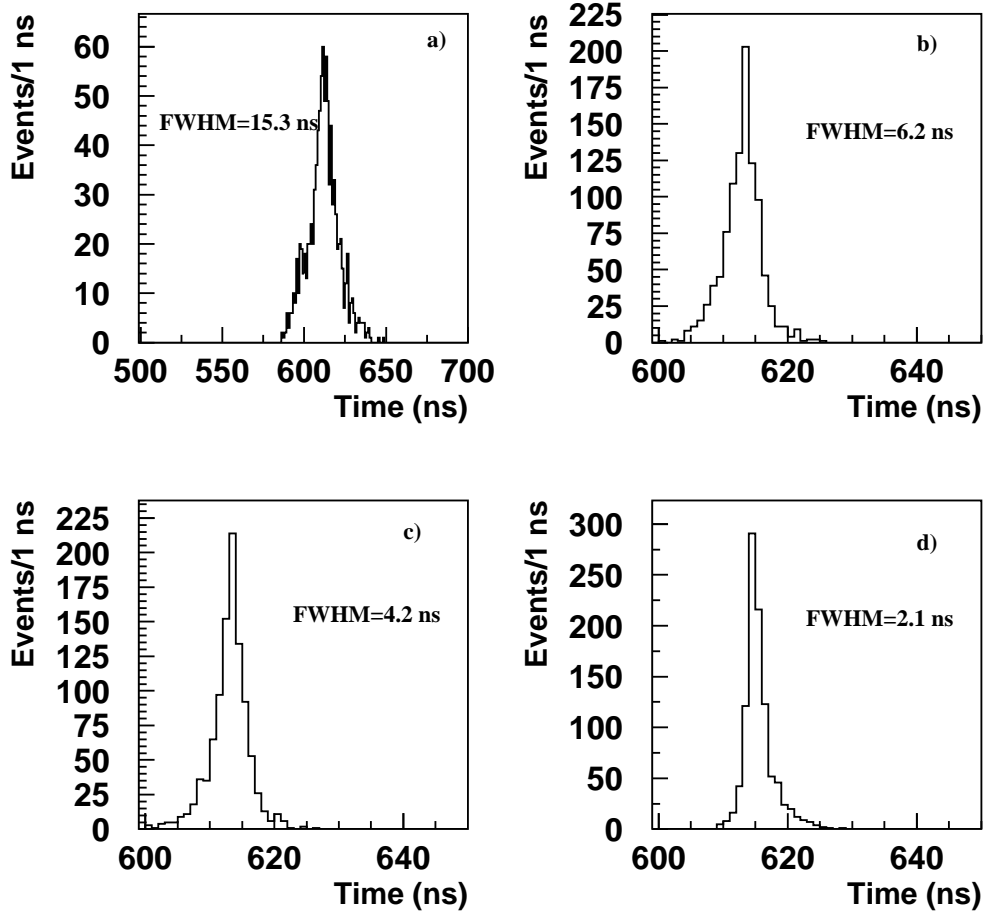,width=140mm,height=140mm}
\end{center}
 \caption{\em Simulated time distributions of bunched positrons
at the MCP for an initial chopped pulse of 300 ns. The cuts on the longitudinal energy
$E_{||}$ of positrons emitted from the moderator are: a) $>0$ eV, b) $>$1.5 eV c) $>$2.0
eV and d)$>$ 2.5 eV. The initial positron energy distribution is taken from the
measurements shown in Fig. \ref{moderator}b, the angular distribution is taken to be
isotropic. }
\label{time}
\end{figure}

In Fig. \ref{time} simulated time distributions of bunched positrons at the target are
shown for different cuts on the longitudinal kinetic energy $E_{||}$ of positrons emitted
from the moderator. The best time resolution with the new bunching method is about 2.1 ns
(FWHM) for an initial pulse duration of 300 ns. It is also seen that the shape and width
of the distributions are affected by the cut, i.e. by the degree to which the moderated
positrons are mono-energetic, as expected from Liouville's theorem.

\section{Experimental results}

\subsection{Moderated positrons}

It is well known that the spectrum of the longitudinal energy of moderated positrons
strongly depends on the quality of the moderator and can be significantly improved by
annealing the W-film in situ, see e.g. \cite{schultz}. The moderator annealing in our
setup was performed at 2000$^o$C by bombarding the foil with an electron beam ( $\simeq$
25 W, 10 kV) for about 10 minutes in a vacuum of $\approx 10^{-8}$ mbar.

\begin{figure}
\begin{center}
  \epsfig{file=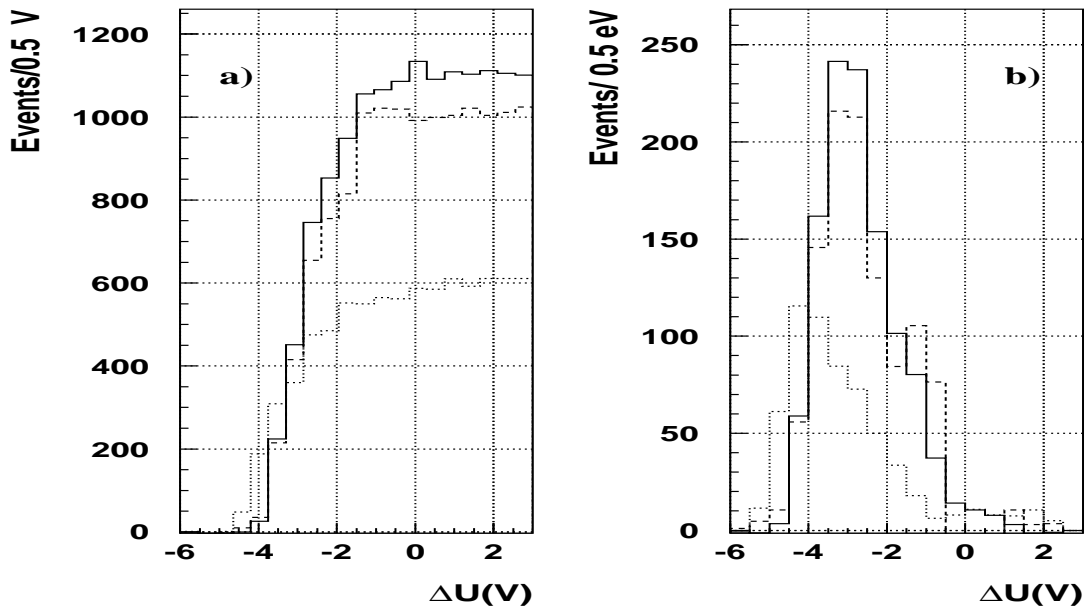,width=150mm,height=90mm}
\end{center}
 \caption{\em a) Positron yield as a function of the
 potential difference between the moderator and the chopper grid.
 b) longitudinal kinetic energy distribution of moderated
 positrons for the W(100) single crystal, moderated before (dotted),
$\sim 1$ h after (solid), and two days after(dashed) in situ annealing.}
\label{moderator}
\end{figure}

Fig. \ref{moderator}a shows the DC positron intensity as a function of the potential
difference $\Delta U$  between the moderator and the chopper grid. The longitudinal
energy spectra are obtained from the derivatives of the corresponding intensity curves
and are shown in Fig. \ref{moderator}b. The spectra are taken before, $\sim 1$ h after
and two days after the annealing. It is seen that the positron yield increases due to
annealing almost by a factor two. The FWHM of the energy distributions also changes from
$\simeq$ 3 eV obtained before to $\simeq$ 2 eV measured after annealing.  The spectra
taken two days after annealing illustrate degradation of the moderator surface through
interactions with a residual gas, which results in a broadening of the energy spectrum
and a more isotropic re-emission of the positrons, i.e. in an increase of the beam
emittance.

\subsection{Positron bunch width}

\begin{figure}
\begin{center}
  \epsfig{file=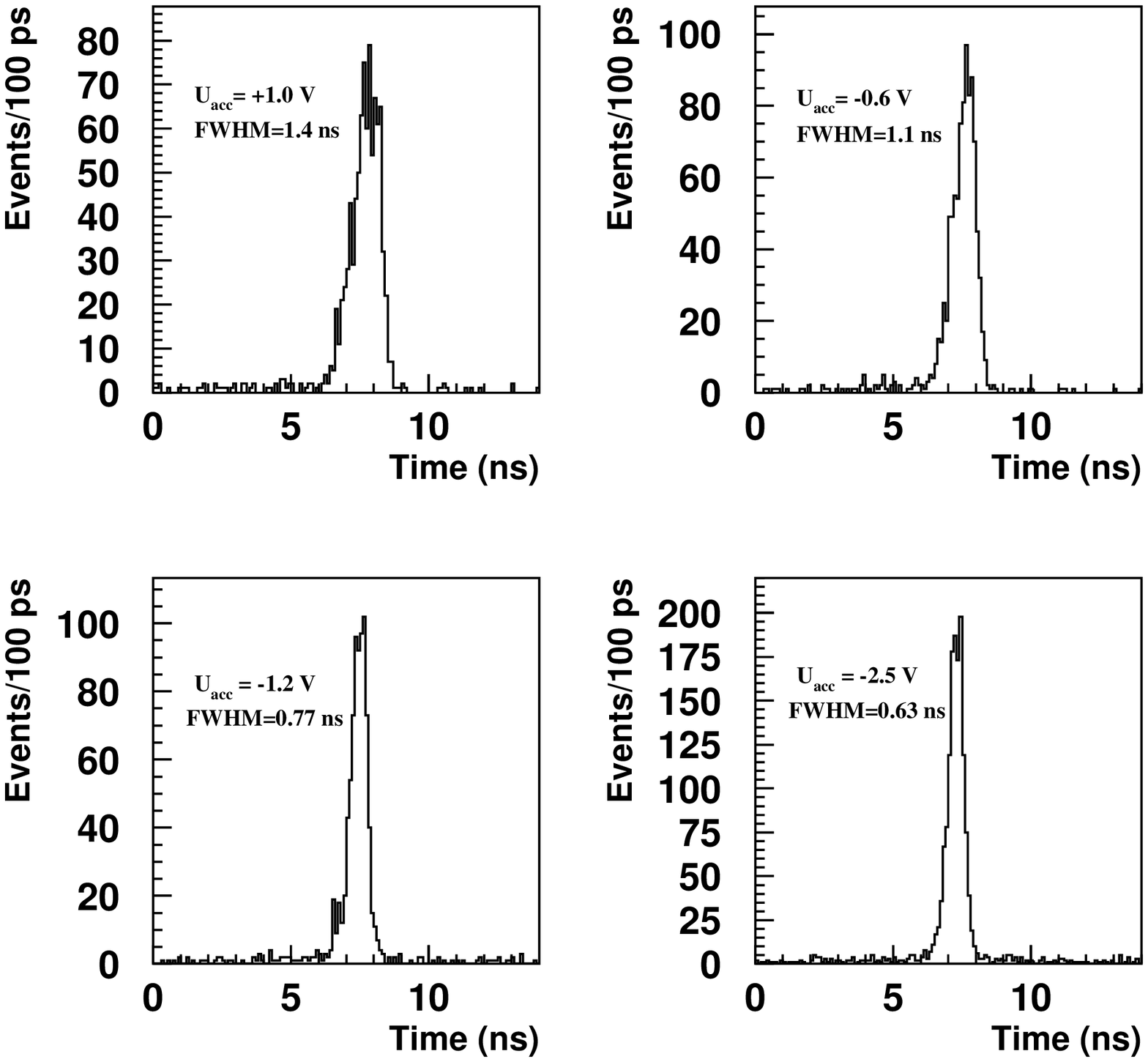,width=140mm,height=140mm}
\end{center}
 \caption{\em Time distribution of pulsed positrons
at the target position, measured for an initial positron pulse of 90 ns and for different
potential differences $\Delta U$ between the moderator and the grid, indicated on the
plots.} \label{t_moder}
\end{figure}

In Fig. \ref{t_moder} the measured time distributions of pulsed positrons at the target
for are shown for an initial pulse of 90 ns and for different values of the retarding
voltage between the moderator and the chopper grid. It is seen that the FWHM of the
distributions is very sensitive to the energy spread $\Delta E$ of the moderated
positrons changing from 1.4 ns to 0.63 ns, for $\Delta E \sim 5$ V ($\Delta U=+1$ V) and
$\Delta E\sim 1.5$ V ($\Delta U=-2.5$ V), respectively. It should be noted, that the
measurements shown in Figure \ref{moderator} were performed by the retarding potential
method, using  the grid  as an energy analyzer. However, due to the inhomogeneity of the
electric field formed by the grid, this method probably does not have an energy
resolution better than $\sim 1$ eV and a significant contribution to the FWHM of the
positron energy spectra can be expected.


The measurement results demonstrate that the degree, to which the positrons emitted from
the moderator are mono-energetic, is an important parameter. Thus, it is crucial to have
a well-conditioned stable moderator in order to get a high performance of the beam. This
observation is in qualitative agreement with the results of simulations. \footnote{The
detail comparison of beam simulations and measurement results will be reported
elsewhere.}

\begin{figure}
\begin{center}
  \epsfig{file=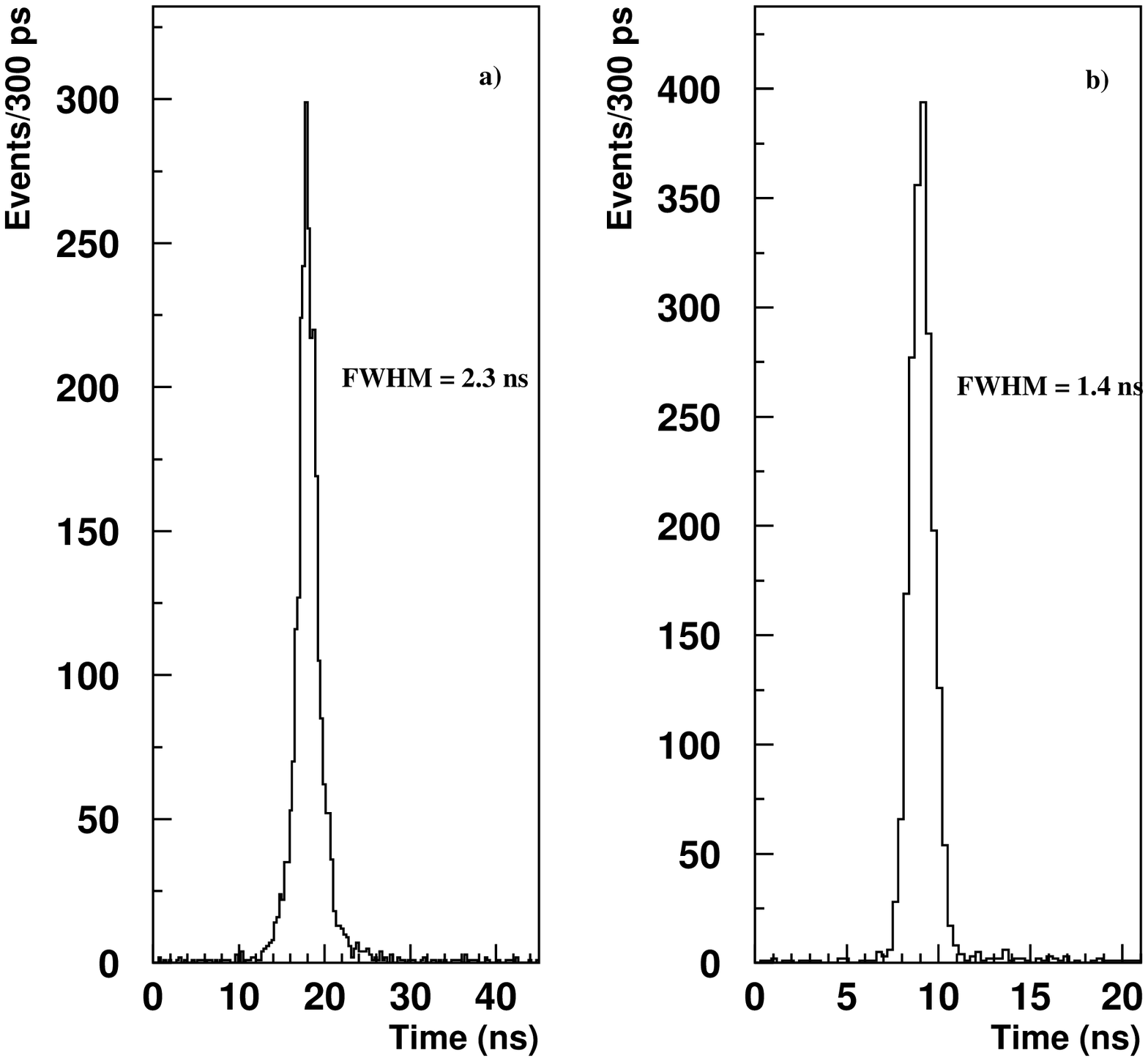,width=140mm,height=100mm}
\end{center}
 \caption{\em Measured time distribution of pulsed positrons at the target
 position for an initial pulse duration of a) 300 ns and b) 120 ns.}
\label{250ns}
\end{figure}

In Fig. \ref{250ns}a, b the measured time distributions of positrons arriving at the MCP
are shown for initial positron pulse durations of 300 ns and 120 ns, respectively. The
FWHM of the distributions are estimated to be 2.3 and 1.4 ns. These values are comparable
with those expected from Monte Carlo simulations, see e.g. Fig. \ref{time}d. The
compression ratio is $\simeq 100$, which is a factor 5 better than the values previously
obtained by Iijima et al. \cite{iijima}, reporting a compression from 50 ns to  2.2 ns,
and by Tashiro et al. \cite{tashiro}, reporting a compression from 30 ns to 1.4 ns,
respectively. For the bunch width of $\sim 2.3$ ns and a repetition period of $ 1\mu s$
our pulsing efficiency is $\simeq$ 31 \%, which is also a factor 6 better than reported
in \cite{iijima}.

\begin{figure}
\begin{center}
  \epsfig{file=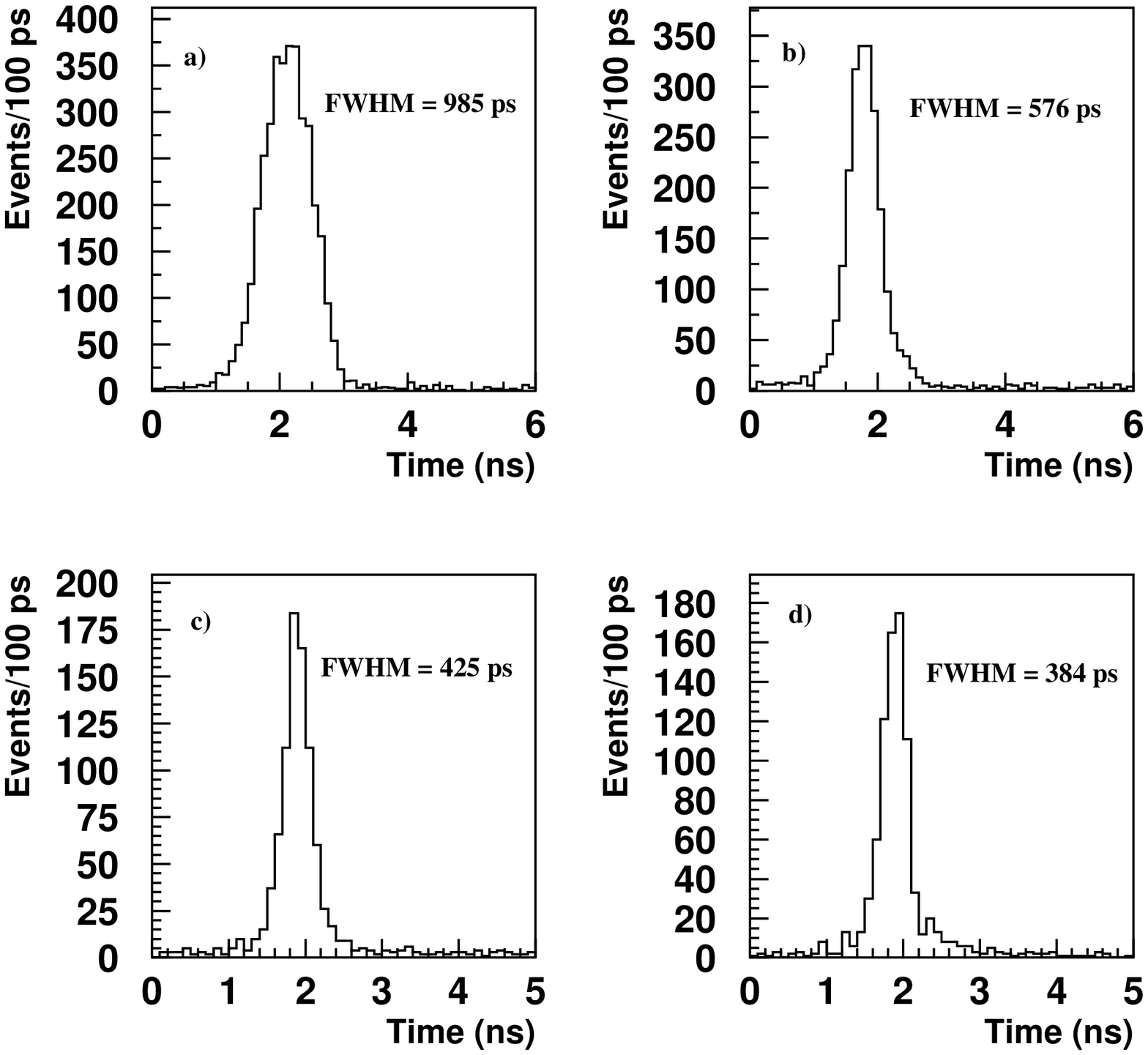,width=140mm,height=140mm}
\end{center}
\vspace{1.0cm}
 \caption{\em  Time distribution of bunched positrons at the target
for a bunch pulse width corresponding to an initial positron pulse of 90 ns and chopper
pulse durations of a) 80 ns, b) 70 ns, c) 60 ns and d) 50 ns, respectively. See text for
details.} \label{90ns}
\end{figure}

It should be noted, that the peak to (flat) background ratio ($\simeq 10^2$) of the
distributions shown in Fig.\ref{250ns}a,b is in agreement with expectations from the
accidental coincidences of START and STOP signals from i) two different positrons, mostly
due to the MCP detection inefficiency, or ii) from non-positron related events due to the
MCP noise. However, the non-Gaussian tails of the distributions in Fig.\ref{250ns}a,b
(see also Fig. \ref{90ns}) are slightly worse than expected from simulations. There are
several contributions to these tails due to: i) the angular distribution of the moderated
positrons and the fact that they are not mono-energetic, ii) the extraction of slow
positrons from the moderator, iii) deviations of the pulse applied to the bunching
electrodes from the calculated ideal shape, iv) heterogeneous electric and magnetic
fields, v) time jitter of the detecting electronics, etc.

\section{Possible applications of the pulsed beam}

It is well known that positron annihilation lifetime spectroscopy (PALS) based on intense
pulsed positron beams with a bunch width shorter than $\simeq$ 1 ns is useful for many
interesting applications, such as studies of polymers, coatings, measurements of porosity
of thin films,  etc \cite{jean}. In these studies information on a sample structure is
extracted from the results of the lifetime measurement of $ o-Ps$ formed inside the
sample. The vacuum value of the $o-Ps$ lifetime is shortened by collisional pick-off
annihilation and ranges typically from a fraction of a ns to tens of ns, see e.g.
\cite{jean}. For high quality PALS spectra measurements the important characteristics of
the positron bunch at the sample are: i) the bunch width, typically $ < 1$ ns, ii) peak
to background ratio, typically $\geq 10^2$; this is an important factor for accurate
measurements of low-intensity $o-Ps$ components, and iii) shape of the time profile
(resolution function), typically one or two Gaussians with small tails. This is important
for measurements of short $o-Ps$ lifetimes.

The flexibility of the developed pulsing scheme encourages us to test the generation of
positron pulses with a width  $< 1$ ns. To avoid the influence of significant bunching
pulse distortions on the positron time profile at the target, we try to eliminate them by
using only a part of the bunching pulse, where the aberrations of the shape are within a
few \% (typically, the aberrations were significant at the beginning and at the end of
the pulse). For this purpose the initial positron pulse was shortened to a desired
duration by the chopper pulse in such a way, that the chopped positrons passing the
bunching electrode received the correct bunching voltage from the selected (central) part
of the buncher pulse, where aberration effects are negligible.

 \begin{figure}
\begin{center}
  \epsfig{file=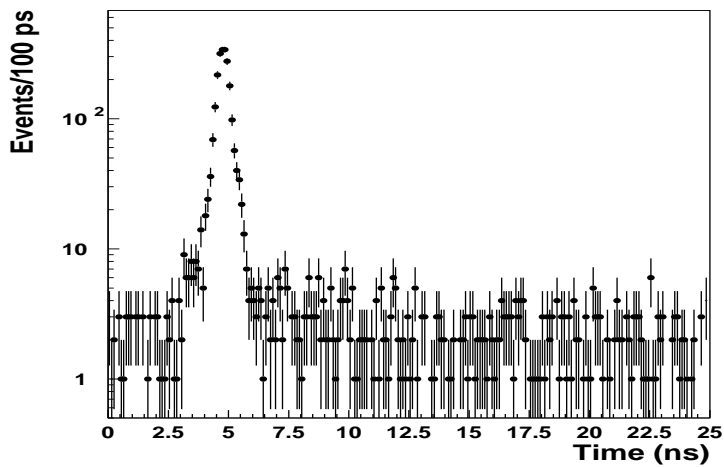,width=100mm,height=70mm}
\end{center}
\vspace{1.0cm}
 \caption{\em Time distribution of positrons at the target with
  a FWHM of 576 ps.}
\label{500ps}
\end{figure}

In Fig. \ref{90ns} the results of these tests are shown for different chopper pulse
durations, varying from 80 ns to 50 ns and a bunch pulse width corresponding to an
initial positron pulse of 90 ns. It is seen that a positron bunch width as short as
$\simeq $ 400 ps can be obtained with this pulsing method. In Fig. \ref{500ps} the time
distribution of Fig.\ref{90ns}b is shown with a logarithmic scale for more details.
Although it has low statistics (because of the small $^{22}$ Na source activity the
counting rate is just $\sim 1$ Hz) the FWHM and the peak to background ratio ($\sim
10^2$) are compatible with the pulsed beams for application to polymer films, see e.g.
Ref.\cite{tashiro}. \footnote{We assume that the convolution of this spectrum with a time
resolution of $\simeq$ 200-300 ps of a fast BaF$_2$ $\gamma-$detector, usually used for
PALS measurements, will not significantly affect the FWHM of the spectrum.}

We think that further increasing the statistics with a higher activity source, e.g. using
the radioisotope $^{18}$F \cite{ulisse}, and an improvement of the time profile of the
bunched positrons results in a suitable beam for applications on thin film measurements.

\section{Summary}

We have developed a high-efficiency pulsed slow positron beam for experiments with
orthopositronium in vacuum. The new pulsing scheme is based on a double-gap coaxial
buncher powered by an RF pulse of a  specially designed shape, which is produced by an
arbitrary waveform generator. With the modulation of the positron velocity in two gaps,
their time-of-flight to a target is adjusted. This pulsing scheme allows to minimize
non-linear aberrations in the bunching process and to achieve a compression ratio limited
mainly by the intrinsic energy spread of the initial positrons. The flexibility of the
new scheme allows us to efficiently compress the positron pulse with an initial pulse
duration ranging from $\sim$ 300 to 50 ns into a bunch of  2.3  to  0.4 ns width,
respectively. A compression ratio of $\simeq 100$ and a pulsing efficiency of $\simeq
30$\% were achieved for a repetition period of 1~$\mu s$, which is 5 to 6 times better
than previously reported numbers.

Both, simulation and measurement results, demonstrate that i) the degree to which the
positrons emitted from the moderator are mono-energetic, and ii) the precision of the
bunching pulse waveform are important for a high performance of the beam. This will
require the possible construction of a new, well-conditioned moderator with narrow ($\leq
1$ eV) londitudinal energy spread of moderated positrons. In general, the developed beam
is suitable for experiments with $o-Ps$ in vacuum, mentioned in section 1.

Preliminary results on the generation of short positron bunches for PALS applications are
encouraging. However, further work to increase the beam intensity and possibly to improve
the time profile of bunched positrons is required.

{\bf Acknowledgments}

We thank ETH Z\"urich, the Swiss National Science Foundation, and the INR Moscow for
support given to this research. We acknowledge the French institutions which have
supported the project: Region Rh\^one-Alpes through an "Avenir project", the French
Ministery of Foreign Affairs through an ECONET and a PAI program. We would like to thank
A. Gonidec, J-P. Peigneux, V. Postoev, P. Nedelec, V. Samoylenko for support and help in
this work.
   The crucial assistance of L. Knecht,  L. Kurchaninov, A. Turbabin,
 A. Shnyrev and L. Zacharov
in the design and construction of the beam and electronics is  gratefully acknowledged.

\end{document}